\documentstyle[twocolumn,prl,aps,psfig]{revtex}

\begin{document}
\title{High-sensitivity optical measurement of mechanical Brownian motion}
\author{Y.\ Hadjar, P.F.\ Cohadon, C.G.\ Aminoff\thanks{%
Helsinki University of Technology, Department of Engineering Physics and
Mathematics, 02015\ HUT\ Espoo, Finland}, M.\ Pinard, A.\ Heidmann\thanks{%
e-mail : heidmann@spectro.jussieu.fr}}
\address{Laboratoire Kastler Brossel\thanks{%
Laboratoire de l'Universit\'{e} Pierre et Marie Curie et de l'Ecole Normale
Sup\'{e}rieure associ\'{e} au Centre National de la Recherche Scientifique},
Case 74, 4 place Jussieu, F75252\ Paris Cedex 05, France}
\maketitle

\begin{abstract}
We describe an experiment in which a laser beam is sent into a high-finesse
optical cavity with a mirror coated on a mechanical resonator.\ We show that
the reflected light is very sensitive to small mirror displacements.\ We
have observed the Brownian motion of the resonator with a very high
sensitivity corresponding to a minimum observable displacement of $2\times
10^{-19}$ ${\rm m}/\sqrt{{\rm Hz}}$.
\end{abstract}

{\bf PACS :} 05.40.Jc, 04.80.Nn, 42.50.Lc\bigskip

Thermal noise plays an important role in many precision measurements \cite
{Saulson90}. For example, the sensitivity in interferometric
gravitational-wave detectors \cite{Bradaschia90,Abramovici92,Abramovici96}
is limited by the Brownian motion of the suspended mirrors which can be
decomposed into suspension and internal thermal noises. The latter is due to
thermally induced deformations of the mirror surface. Experimental
observation of this noise is of particular interest since its theoretical
evaluation strongly depends on the mirror shape and on the spatial matching
between light and internal acoustic modes \cite
{Bondu95,Gillespie95,Nakagawa97,Levin98,Bondu98}. It is also related to the
mechanical dissipation mechanisms which are not well known in solids \cite
{Startin98}.\ Mirror displacements induced by thermal noise are however very
small and a highly sensitive displacement sensor is needed to perform such
an observation.

Monitoring extremely small displacements has thus become an important issue
in precision measurements \cite{Braginsky93b} and several sensors have been
developed.\ A technique commonly used for the detection of gravitational
waves by Weber bars is based on capacitive sensors \cite{Bocko96}.\ Another
promising technique consists in optical transducers \cite{Mio95,Tittonen99}.
Reflexion of light by a high-finesse Fabry-Perot cavity is very sensitive to
changes in the cavity length.\ Such a device can thus be used to monitor
displacements of one mirror of the cavity, as it has been proposed for
gravitational wave bar detectors where the mirror is mechanically coupled to
the bar \cite{Richard92,Conti98}, or for the detection of Brownian motion in
gravitational wave interferometers \cite{Stephens93,Bernardini98}. In this
letter we report a high-sensitivity observation of the Brownian motion of
internal modes of a mirror. The sensitivity reached in our experiment is
better than that of present sensors and comparable to the one expected in
gravitational wave interferometers.

We use a single-ended Fabry-Perot cavity composed of an input coupling
mirror and a totally reflecting back mirror. The intracavity intensity shows
an Airy peak when the cavity length is scanned through a resonance, and the
phase of the reflected field is shifted by $\pi$. The slope of this phase
shift strongly depends on the cavity finesse and for a lossless resonant
cavity, a displacement $\delta x$ of the back mirror induces a phase shift $%
\delta \varphi _{x}$ of the reflected field on the order of 
\begin{equation}
\delta \varphi _{x}\simeq 8{\cal F}\frac{\delta x}{\lambda },
\label{Eq_dPhiOut}
\end{equation}
where ${\cal F}$ is the cavity finesse and $\lambda $ is the optical
wavelength.\ This signal is superimposed to the phase noise of the reflected
field.\ If all technical noise sources are suppressed, the phase noise $%
\delta \varphi _{n}$ corresponds to the shot noise of the incident beam 
\begin{equation}
\delta \varphi _{n}\simeq \frac{1}{2\sqrt{\overline{I}}},  \label{Eq_dPhiIn}
\end{equation}
where $\overline{I}$ is the mean incident intensity counted as the number of
photons per second. The sensitivity of the measurement is given by the
minimum displacement $\delta x_{min}$ that yields a signal of the same order
of magnitude as the noise 
\begin{equation}
\delta x_{min}\simeq \frac{\lambda }{16{\cal F}\sqrt{\overline{I}}}.
\label{Eq_dxmin}
\end{equation}
One expects to be able to detect a displacement corresponding to a small
fraction of the optical wavelength for a high-finesse cavity and an intense
incident beam.

In our experiment the coupling mirror has a curvature radius of 1 meter and
a typical transmission of 50 ppm (Newport high-finesse SuperMirror). The
back mirror is coated on the plane side of a small plano-convex mechanical
resonator made of silica. The coating has been made at the {\it Institut de
Physique Nucl\'{e}aire (Lyon)} on a 1.5-mm thick substrate with a diameter
of 14 mm and a curvature radius of the convex side of 100 mm. The two
mirrors are mounted in a rigid cylinder which defines the distance and the
parallelism between them.\ The cavity length is close to 1 mm so that the TEM%
$_{00}$ optical mode of the cavity has its waist in front of the back mirror
with a size of 90 $\mu $m.

The mirror motion is due to the excitation of internal acoustic modes which
have been extensively studied for plano-convex resonators \cite
{Wilson74,Stevens86,Heidmann97,Pinard99}. For a curvature radius of the
convex side much larger than the thickness of the resonator, those modes can
be described as gaussian modes confined around the central axis of the
resonator.\ The intracavity field experiences a phase shift proportional to
the longitudinal deformation of the resonator averaged over the beam waist 
\cite{Bondu95,Gillespie95,Nakagawa97,Levin98,Bondu98,Pinard99} and only
compression modes which induce such a longitudinal deformation are coupled
with the light.\ In the following we focus on the fundamental mode which has
a waist equal to 3.4 mm and a resonance frequency close to 2 MHz.

We have measured the optical characteristics of the cavity.\ Its bandwidth
is equal to 1.9 MHz and its free spectral range is equal to 141 GHz. These
values correspond to a cavity length of 1.06 mm and a finesse ${\cal F}$ of
37000. We also measured the reflexion coefficient of the cavity at resonance
to derive the transmission of the coupling mirror and the cavity losses.\ We
found a transmission of 60 ppm and losses equal to 109 ppm.

\begin{figure}
\centerline{\psfig{figure=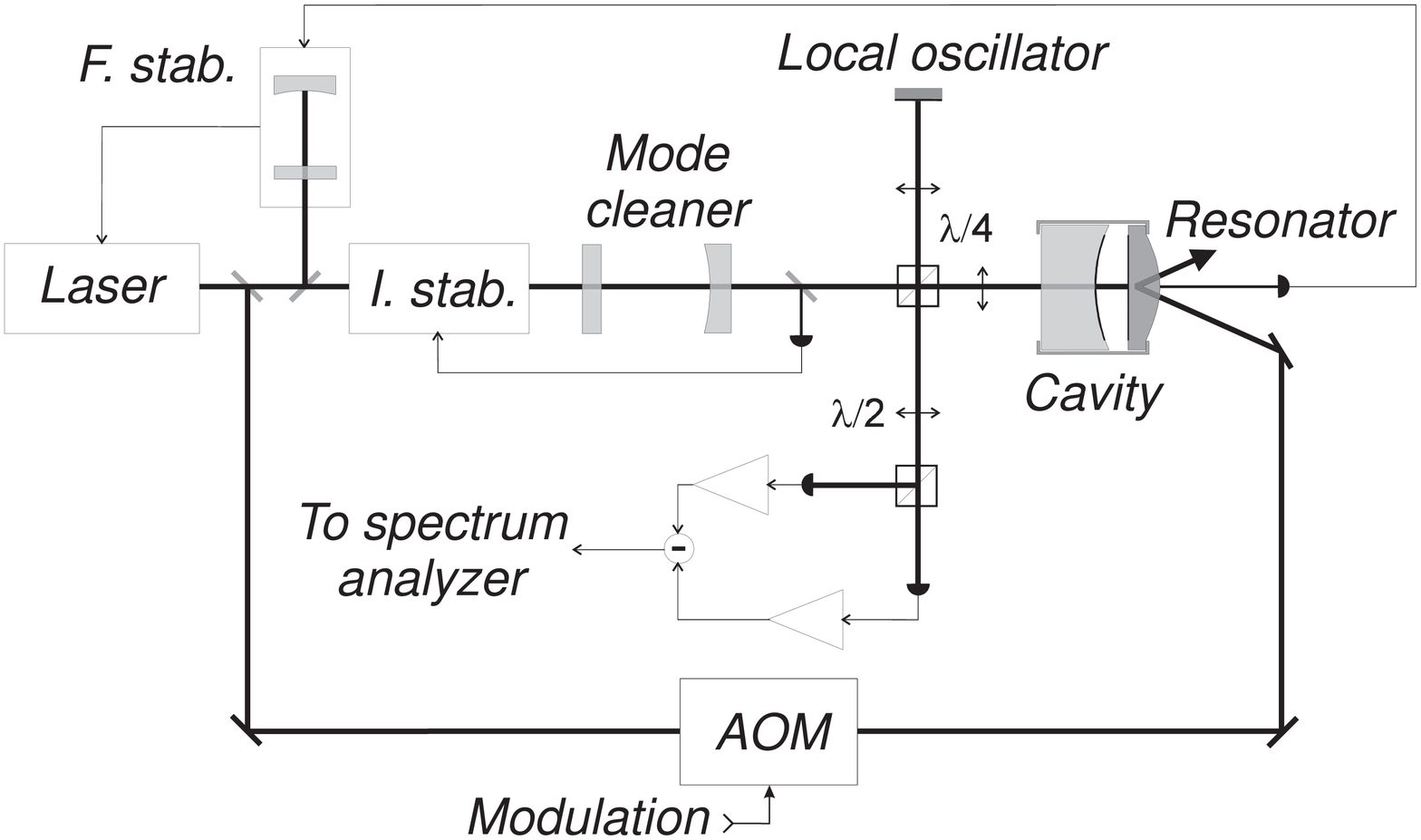,width=7.5cm}}
\vspace{2mm}
\caption{Experimental setup. A light beam supplied by a frequency (F. stab.)
and intensity (I. stab.) stabilized titane-sapphire laser is sent into a high-finesse cavity
composed of a coupling mirror and a highly-reflecting back mirror coated on a mechanical
resonator. Phase fluctuations of the reflected field are measured by
homodyne detection. An auxiliary beam with modulated intensity (AOM) is used to optically
excite the acoustic modes of the resonator}
\label{Fig_Setup}
\end{figure}

The light entering the cavity is supplied by a titane-sapphire laser working
at 810 nm and frequency-locked to a stable external cavity by sideband
techniques \cite{Drever83}. We use a triple servoloop to monitor the laser
frequency via a mirror mounted on a piezoelectric ceramic and an
electro-optic modulator placed inside the laser cavity.\ The residual jitter
is mainly concentrated at low frequency and is less than 3 kHz rms. The
frequency noise is less than 15 ${\rm mHz}/\sqrt{{\rm Hz}}$ above 1 MHz. The
laser frequency is locked to a resonance of the high-finesse cavity by
monitoring the residual light transmitted by the back mirror via a control
of the external cavity length. We use a mode cleaner to reduce the
astigmatism of the laser beam.\ It consists of a non-degenerate linear
cavity locked at resonance with the laser.\ The transmitted beam corresponds
to a fundamental mode of the cavity which is gaussian TEM$_{00}$. The mode
matching of the resulting beam with the high-finesse cavity is equal to
98\%. The intensity after the mode cleaner is actively stabilized by a
variable attenuator inserted in front of the mode cleaner.\ One gets a 100-$%
\mu $W incident power on the high-finesse cavity with residual relative
intensity fluctuations less than 0.2\%. Note that the incident power is low
enough to neglect quantum effects of radiation pressure\cite
{Tittonen99,Heidmann97,Pinard99}.\ Quantum noise induced by radiation
pressure is less than 1\% of the phase noise $\delta \varphi _{n}$.

The phase of the field reflected by the high-finesse cavity is measured by
homodyne detection (fig. \ref{Fig_Setup}). The reflected field is mixed on
two photodiodes (FND100 from EGG Instruments) with a 10-mW local oscillator
derived from the incident beam. We use a set of quarter-wave plates, a
half-wave plate and polarizing beamsplitters to separate and mix those
fields.\ The two photocurrents are preamplified with wideband and low-noise
transimpedance amplifiers and their difference is sent to a spectrum
analyzer. The overall quantum efficiency of the detection system is equal to
91\%.\ The signal obtained on the spectrum analyzer is proportional to the
fluctuations of the quadrature component of the reflected field in phase
with the local oscillator. A servoloop monitors the length of the local
oscillator arm so that we detect the phase quadrature of the reflected
field. This setup is thus similar to an interferometer with dissymmetric
arms.\ It indeed performs an interferometric measurement of the back-mirror
position, the sensitivity being increased by the cavity finesse.

The last part of the experimental setup is used to optically excite the
mechanical resonator.\ A 500-mW auxiliary beam derived from the
titane-sapphire laser is intensity-modulated by an acousto-optic modulator
and reflected from the rear on the back mirror.\ A modulated radiation
pressure force is thus applied to the resonator.\ The amplitude of this
force can be changed by varying the depth of the intensity modulation.\ The
auxiliary laser beam is uncoupled from the cavity by frequency filtering due
to an optical frequency shift of 200 MHz induced by the acousto-optic
modulator and by spatial filtering due to a tilt angle of 10$%
{{}^\circ}%
$ between the beam and cavity axes. We have checked that the auxiliary beam
has no spurious effect on the homodyne detection.

\begin{figure}
\centerline{\psfig{figure=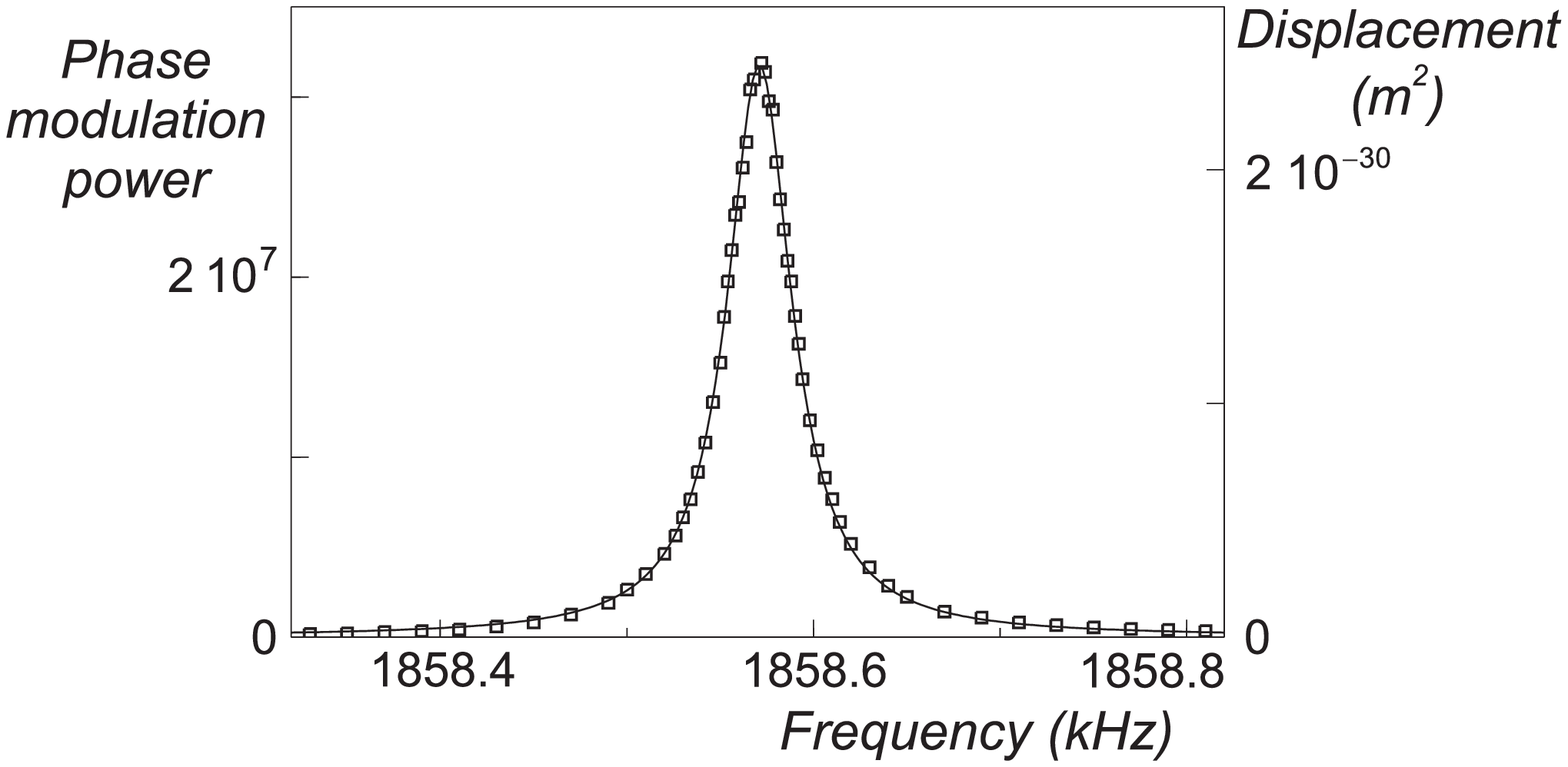,width=7.5cm}}
\vspace{2mm}
\caption{Mechanical response of the resonator. Squares represent the power of
phase modulation of the reflected field normalized to the shot-noise level.
Each square corresponds to a different modulation frequency of the optical excitation,
around the fundamental resonance frequency of the resonator. The solid curve is a
lorentzian fit of the resonance. Vertical scale on the right represents the equivalent
displacement in $m^{2}$}
\label{Fig_Modul}
\end{figure}

Figure \ref{Fig_Modul} shows the experimental result of the optical
excitation.\ Each square is obtained for a different modulation frequency of
the auxiliary laser beam around the expected frequency for the fundamental
mode of the mechanical resonator. The power of phase modulation of the
reflected field is normalized to the shot-noise level, independently
measured by sending only the local oscillator in the homodyne detection.\ We
have checked that the phase noise of the reflected field corresponds to the
shot-noise level when the laser is out of resonance with the high-finesse
cavity.\ Any deviation of the phase from the shot-noise level is thus due to
the interaction of the light with the cavity. Such a deviation reflects the
mirror motion and\ the resonance in figure \ref{Fig_Modul} corresponds to
the excitation of the fundamental acoustic mode of the resonator.\ The solid
curve is a lorentzian fit which shows that the mechanical response has a
harmonic behavior around the resonance frequency with a quality factor $Q$
of 44000.

As explained in the end of this paper we have calibrated the measured
displacement and the resulting scale is shown on the right of figure \ref
{Fig_Modul}.\ The displacement at resonance corresponds to an amplitude of $%
1.6\times 10^{-15}$ m.\ One can estimate the radiation pressure exerted by
the auxiliary beam as $F_{rad}=2\hbar k\delta I=1.2\times 10^{-9}$ N where $%
2\hbar k$ is the momentum exchange during a photon reflection and $\delta I$
is the intensity modulation.\ One thus finds that the mechanical
susceptibility $\chi \left[ \Omega \right] $ has a lorentzian shape around
the mechanical resonance frequency $\Omega _{M}$%
\begin{equation}
\chi \left[ \Omega \right] =\frac{\chi _{0}}{1-\Omega ^{2}/\Omega
_{M}^{2}-i/Q},  \label{Eq_Chi}
\end{equation}
with $\chi _{0}=3.2\times 10^{-11}$ ${\rm m}/{\rm N}$.

\begin{figure}
\centerline{\psfig{figure=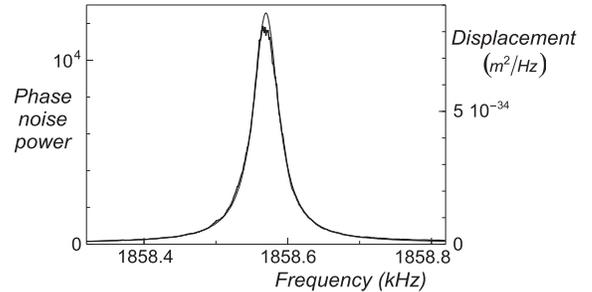,width=7.5cm}}
\vspace{2mm}
\caption{Phase noise spectrum of the reflected field normalized to the shot-noise level
for a frequency span of 500 Hz around the fundamental resonance frequency of the
resonator. The peak reflects the Brownian motion of the resonator at room
temperature. The thin line is a theoretical estimation of the thermal noise. Vertical scale
on the right represents the equivalent displacement in ${\rm m}^{2}/{\rm Hz}$}
\label{Fig_Thermal}
\end{figure}

Figure \ref{Fig_Thermal} shows the phase noise spectrum of the reflected
beam obtained with a resolution bandwidth of 1 Hz and for the same frequency
range (500-Hz span around the fundamental resonance frequency). The
auxiliary laser beam is now turned off (no optical excitation) and the
resonator is at room temperature. The spectrum is obtained by an average
over 1000 scans of the spectrum analyzer. It is normalized to the shot-noise
level and the vertical scale is smaller than the one of figure \ref
{Fig_Modul}. The thin line in figure \ref{Fig_Thermal} corresponds to a
theoretical estimation of the thermal noise at 300 K by\ using the
mechanical susceptibility $\chi \left[ \Omega \right] $ derived from optical
excitation (eq. \ref{Eq_Chi}).\ Note that there is no adjustable parameter
and the excellent agreement with experimental data clearly shows that the
peak observed in figure \ref{Fig_Thermal} corresponds to the thermal noise
of the fundamental mode of the resonator.

We have calibrated the observed displacements by frequency modulation of the
incident laser beam.\ The detuning between the laser and the cavity
resonance indeed only depends on the optical frequency and on the cavity
length.\ A displacement $\delta x$ of the back mirror is thus equivalent to
a frequency modulation $\delta \nu $ of the laser related to $\delta x$ by 
\begin{equation}
\frac{\delta \nu }{\nu }=\frac{\delta x}{L},  \label{Eq_dvdx}
\end{equation}
where $\nu $ is the optical frequency and $L$ the cavity length. We can thus
calibrate the observed displacements by measuring the frequency modulation
which yields the same phase signal for the reflected field.

The frequency modulation of the laser beam is obtained by applying a
sinusoidal voltage on the internal electro-optic modulator of the laser. We
determine the amplitude $\delta \nu $ of modulation by locking the mode
cleaner at half-transmission and by measuring the intensity modulation of
the transmitted beam.\ This intensity modulation is proportional to the
ratio $\delta \nu /\nu _{cav}$ between the amplitude of frequency modulation
and the cavity bandwidth $\nu _{cav}$ of the mode cleaner.\ We have
determined this bandwidth with a good accuracy by measuring the transfer
function of the mode cleaner at resonance for an intensity-modulated
incident beam.

\begin{figure}
\centerline{\psfig{figure=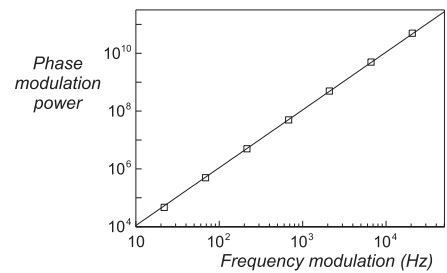,height=3.5cm}}
\vspace{2mm}
\caption{Power of phase modulation of the reflected field normalized to the shot-noise
level as a function of the amplitude of the laser frequency modulation.
The solid line is a linear fit of the data points represented by squares}
\label{Fig_Sensitivity}
\end{figure}

Figure \ref{Fig_Sensitivity} shows the result of the calibration. We applied
a sinusoidal voltage to the laser with different amplitudes at a frequency
of 2 MHz. The horizontal axis represents the amplitude $\delta \nu $ of
frequency modulation determined from the mode-cleaner cavity.\ The vertical
axis corresponds to the power of phase modulation observed in the field
reflected by the high-finesse cavity.\ Experimental results represented by
squares are obtained with a 1-Hz resolution bandwidth of the spectrum
analyzer and are normalized to the shot-noise level. The linear fit (solid
curve in figure \ref{Fig_Sensitivity}) has a slope equal to 2 as expected in
log-log scales since the power of phase modulation must be proportional to
the square of the frequency modulation.\ From equation\ (\ref{Eq_dvdx}) one
can associate a displacement $\delta x$ to any observed phase modulation of
the reflected field. In particular, the shot-noise level corresponds to a
frequency modulation $\delta \nu _{min}$ equal to 96 ${\rm mHz}/\sqrt{{\rm Hz%
}}$. The smallest observable thermal displacement $\delta x_{min}$ which
corresponds to the shot-noise level is thus equal to 
\begin{equation}
\delta x_{min}\left[ 2\ {\rm MHz}\right] =L\frac{\delta \nu _{min}}{\nu }%
=2.8\times 10^{-19}\ {\rm m}/\sqrt{{\rm Hz}}.  \label{Eq_dx2MHz}
\end{equation}

This experimental result can be compared to the theoretical prediction.\
Equation (\ref{Eq_dxmin}) corresponds to a static analysis for a lossless
cavity and for a perfect detection system. Cavity filtering at non zero
frequency and losses reduce the theoretical sensitivity.\ The proper
expression of the minimum displacement at frequency $\Omega $ is 
\begin{equation}
\delta x_{min}\left[ \Omega \right] =\frac{\lambda }{16{\cal F}\sqrt{%
\overline{I}}}\frac{T_{c}+A}{\sqrt{\eta }T_{c}}\sqrt{1+\left( \Omega /\Omega
_{cav}\right) ^{2}},  \label{Eq_dxminTheo}
\end{equation}
where $\eta $ is the quantum efficiency of the detection, $T_{c}$ the
transmission of the coupling mirror, $A$ the cavity losses and $\Omega
_{cav} $ the cavity bandwidth. The cavity behaves like a low-pass filter
with a cutoff frequency $\Omega _{cav}$. We have thus performed another
sensitivity measurement at the frequency of 500 kHz.\ We have found that the
shot-noise level corresponds to a frequency modulation $\delta \nu _{min}$
of 68 ${\rm mHz}/\sqrt{{\rm Hz}}$ and the sensitivity $\delta x_{min}$ is
then equal to 
\begin{equation}
\delta x_{min}\left[ 500\ {\rm kHz}\right] =2\times 10^{-19}\ {\rm m}/\sqrt{%
{\rm Hz}}.  \label{Eq_dx500kHz}
\end{equation}
Both experimental values (eqs.\ \ref{Eq_dx2MHz} and \ref{Eq_dx500kHz}) are
in perfect agreement with theoretical values deduced from equation\ (\ref
{Eq_dxminTheo}) with the parameters of the cavity (finesse ${\cal F}=37000 $%
, coupler transmission $T_{c}=60{\rm \ }$ppm, cavity losses $A=109{\rm \ }$%
ppm, cavity bandwidth $\Omega _{cav}/2\pi =1.9{\rm \ }$MHz, quantum
efficiency $\eta =0.91$, wavelength $\lambda =810{\rm \ }$nm and incident
power $P=\left( hc/\lambda \right) \overline{I}=100$ $\mu $W).\ The
discrepancy is less than 5\%.

In conclusion, we have observed the Brownian motion of internal acoustic
modes of a mirror with a very high sensitivity.\ This result demonstrates
that a high-finesse cavity is a very efficient displacement sensor.\ The
possibility to observe the thermal noise even far on the wings of the
mechanical resonances opens up the way to a quantitative study of the
spectral dependence of the Brownian motion.\ This would allow to
discriminate between different dissipation mechanisms in solids. Let us
emphasize that our device also allows to study with a very high accuracy the
mechanical characteristics of the various acoustic modes (resonance
frequency, quality factor, spatial structure, effective mass) and their
coupling with the light. It is furthermore possible to obtain even larger
sensitivities by increasing the finesse of the cavity or the incident light
power. Mirrors with losses of the order of 1 ppm are now available and
cavity finesses larger than $3\times 10^{5}$ have been obtained \cite
{Rempe92}.\ For an incident power of 1 mW one would obtain a sensitivity
better than $10^{-20}$ ${\rm m}/\sqrt{{\rm Hz}}$.

We gratefully thank J.M.\ Mackowski of the {\it Institut de Physique
Nucl\'{e}aire} ({\it Lyon}) for the optical coating of the mechanical
resonator.\ YH\ acknowledges a fellowship from the {\it Association
Louis de Broglie d'Aide \`{a} la Recherche}.

\end{document}